\documentclass[prb,aps,showpacs,superscriptaddress,floatfix,twocolumn]{revtex4-1}
\usepackage{graphicx}
\usepackage{array}
\usepackage{textcomp}

\begin{document}

\title{Ab-initio investigation of structural, electronic, and optical properties of
(5,0) finite-length carbon nanotube}
\author{Mahdi Tarighi Ahmadpour}
\affiliation{Electroceram research center, Malek Ashtar University of Technology, 
             Shahinshahr, Isfahan, Iran}
\author{S. Javad Hashemifar}
\affiliation{Department of Physics, Isfahan University of Technology, 
             84156-83111 Isfahan, Iran}
\email{hashemifar@cc.iut.ac.ir}
\author{Ali Rostamnejadi}
\affiliation{Electroceram research center, Malek Ashtar University of Technology, 
             Shahinshahr, Isfahan, Iran}
\date{\today} 

\newcommand{\etal}{{\em et al.}}
\begin{abstract}
We use density functional computations to study the size effects
on the structural, electronic, magnetic, and optical properties 
of (5,0) finite carbon nanotubes (FCNT), with length in the range
of 4-44\,\AA. It is found that the structural and electronic 
properties of (5,0) FCNTs, in the ground state, converge at a length
of about 30\,\AA, while the excited state properties exhibit long-range
edge effects. We discuss that curvature effects govern the electronic
structure of short (5,0) FCNTs and enhance energy gap of these systems,
in contrast to the known trend in the periodic limit.
It is seen that compensation of curvature effects in two special
small sizes, may give rise to spontaneous magnetization.
The obtained cohesive energies provide some insights into 
the effects of environment on the growth of FCNTs.
The second-order difference of the total energies reveals
an important magic size of about 15\,\AA.
The optical and dynamical magnetic responses of the FCNTs to polarized 
electromagnetic pulses are studied by time dependent density functional 
theory. The obtained results show that the static and dynamic magnetic 
properties mainly come from the edge carbon atoms.
The optical absorption properties are described in term of local field 
effects and characterized by Casida linear response calculations.
\end{abstract}

\maketitle

\section{Introduction}

Recently, finite-length carbon nanotubes (FCNTs) are of 
vast interest due to their potential applications in enhanced
light emitting,\cite{miyauchi:13} spintronics designing,\cite{mananes:08,du:08}
spin-orbit interaction based optoelectronic,\cite{galland:08}
and heterojunction nanoelectronic devices.\cite{chico:96,santos:09}
It has been found that the electronic properties of sub-nanometer
length FCNTs differ substantially from those of periodic CNTs, 
because of the alternation of electronic
structure from one-dimensional to the zero-dimensional characteristic.
\cite{venema:97,zhu:98,rubio:99}
This fact implies that energy states of FCNTs are strictly affected
by the length\cite{jishi:99,montero:10,sun:08}
and the edge structure of the tube.\cite{yumura:04,li:05,wu:09}
For example, while periodic armchair CNTs represent a metallic character,
finite armchair CNTs exhibit an energy gap 
which decreases periodically toward zero by increasing the length.
In the case of zigzag CNTs the situation is different.
On one hand, periodic zigzag CNTs can be either metal or semiconductor, 
depending on their chiral angle.
On the other hand, electronic structure of finite zigzag CNTs 
are dominated by some localized edge states, 
which are evidenced by scanning tunneling spectroscopy of graphite\cite{niimi:06}
and may induce spin-polarization in the system.\cite{okada:03,wu:09}

Thermodynamic considerations suggest that (5,0) CNT 
is the narrowest feasible zigzag carbon nanotube.\cite{lopez:05} 
Because of its high curvature, the electronic properties 
of this nanotube is more complicated than the larger zigzag nanotubes.\cite{zhu:98} 
The well-known $\sigma^\ast$--\,$\pi^\ast$ rehybridization,
as a consequence of this high curvature, forces the zone-folded
semiconducting band gap of periodic (5,0) CNT to close.\cite{blase:94,machon:02}
While periodic structure of this tube is extensively studied,
\cite{spataru:04,marinopoulos:03,machon:02,liu:02,malic:06} 
there is no systematic study about finite size effects on 
the physical properties of this system.
Moreover, first-principles studies on FCNTs are mainly limited
to the ground state properties and understanding finite size effects 
on the excited states properties
of these systems demands further theoretical investigations. 

The dependence of electronic and optical properties of FCNTs
on their geometric features is promising to design flexible and 
sensitive optoelectronic and spintronics devices. 
The specific aim of this study is to investigate 
length-dependent structural, electronic, magnetic and optical
properties of (5,0) FCNTs by using static and time dependent
first-principles calculations.

\section{Methods}

\begin{figure}
\includegraphics*[scale=0.95]{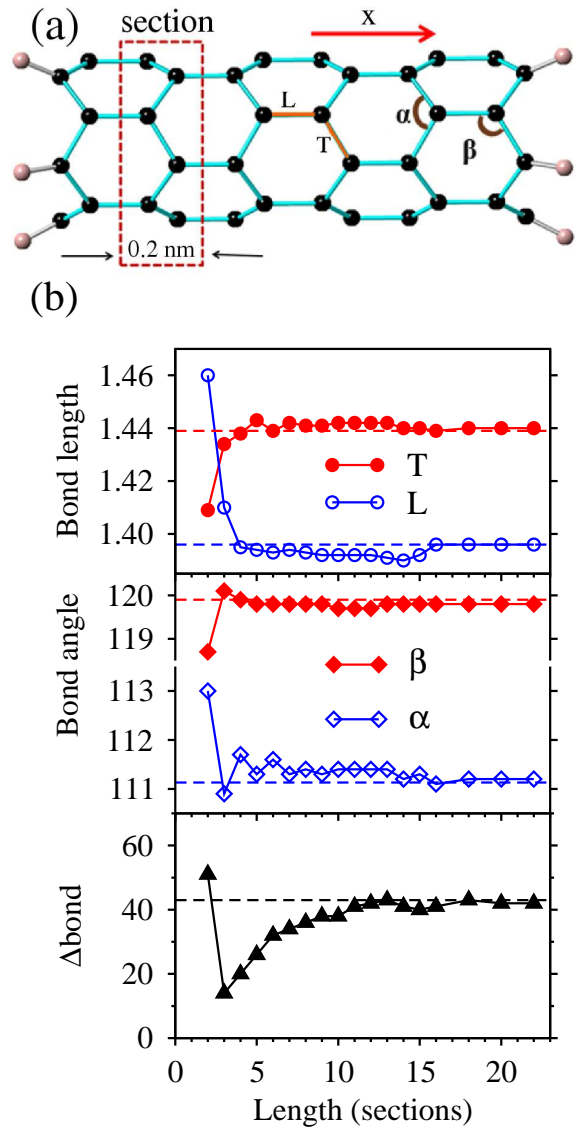}
\includegraphics*[scale=0.95]{fig1-geom}
\caption{\label{geom}
 (a): Schematic diagram of a finite length (5,0) carbon nanotube with 6 sections (S6). 
 The dashed rectangle encloses a tube section. 
 The longitudinal and transverse bonds are indicated by L and T, 
 while $\alpha$ and $\beta$ shows two investigated bond angles. 
 (b): Variation of the L and T bond lengths ($\AA$) and $\alpha$ and $\beta$ 
 bond angles (degree) versus the FCNT's length. 
 The corresponding values for periodic nanotube (infinite-length) 
 are indicated with dashed lines.
 Difference ($\Delta$bond) between the averaged L and T bond lengths 
 over the entire of the tubes. 
}
\end{figure}

The ground state calculations were performed within the Kohn-Sham 
density functional theory (DFT) at the scalar relativistic limit, 
by using the local density (LDA)\cite{perdew:81} 
and local spin density approximations (LSDA) implemented in 
the full-potential \verb+FHI-aims+ code.\cite{blum:09} 
The studied finite nanotubes were terminated by sufficient number of 
hydrogen atoms to saturate their edge dangling bonds. 
Full structural relaxations were applied to the systems 
down to the residual atomic forces of less than 10$^{-3}$ eV/\AA.
The carbon and hydrogen atoms were treated tightly in the calculations. 
The length of the tubes throughout this paper is described in unit 
of a 0.2\,nm thick structural section along the tube axis (Fig.~\ref{geom}a).
The FCNTs are tagged by S${i}$, where $i$ denotes the number 
of sections of the system. 

The optical absorption spectra were evaluated in the framework of 
time dependent DFT (TDDFT) and pseudopotential technique,
implemented in computer package \verb+Octopus+.\cite{andrade:15}
The TDDFT approach has already been used for successful 
interpretation of the excited states properties of a (7,6) FCNT.\cite{tretiak:07}
We used norm-conserving pseudopotentials in the form of Troullier-Martins and 
time-dependent LDA (TD-LDA) kernel.\cite{gross:85} 
It is already discussed that TD-LDA is able to provide reliable results 
for finite systems.\cite{onida:02} 
In \verb+Octopus+, the ground state density is evaluated over a discrete
grid in real space. 
The radius of overlapping spheres around each atom and the grid spacing
were optimized to 3.5 and 0.16\,\AA, respectively. 
Electric dipole response (optical absorption) of the systems are calculated by
explicit real time-propagation of the Kohn-Sham wave functions,\cite{yabana:96} 
in the presence of a weak white uniform electric pulse.
To achieve a fine spectrum, nearly 13000 time-steps are 
taken for a total propagation time of 20\,$\hbar$/eV. 
The frequency response is extracted via the Fourier transform
of the time-dependent dipole moment.
The Casida linear response approach was applied
for characterization of the predicted optical transitions.\cite{casida:11}
In order to evaluate the spin-dipole response of the FCNTs, 
opposite magnetic pulses were applied to the majority and minority spin channels.
\cite{torres:00}

\section{Structural properties}

In this work we investigated the electronic structure and optical 
properties of finite length (5,0) carbon nanotubes composed of up to 22 sections.
The atomic structure of S6, as a prototype, is depicted in Fig.~\ref{geom}a.
The longitudinal and transverse bonds are denoted by L and T,
respectively, and the bond angles are described by $\alpha$ and $\beta$. 
After full structural relaxation, these parameters are averaged 
over the central unit cell of the system and then plotted in Fig.~\ref{geom}b.
It is expected that by increasing the tube length, the edge effects vanish
and central unit cell acts as an infinite length periodic CNT.
According to the figure, the L and T bond lengths, 
after almost 16 sections, converge to the
values of about 1.40 and 1.44\,\AA, respectively, close to 
those reported for periodic (5,0) CNT.\cite{machon:02,liu:02}
For qualitative understanding of the hybridization in the systems,
we compare these bond lengths with those of the ideal 
graphene (1.42\,\AA) and diamond (1.54\,\AA).\cite{saito:98} 

The converged value of the longitudinal bond (1.40\,\AA)
is smaller than the carbon-carbon bond in ideal graphene,
while the converged value of the transverse bond (1.44\,\AA) has slightly 
increased toward the interatomic bond length in diamond.
Hence, it seems that L bond has mainly a graphene like sp$^2$ hybridization,
while T bond may involve a degree of sp$^3$ contribution.

The trends of $\alpha$ and $\beta$ bond angles also confirm the
presence of a remarkable sp$^3$ hybridization in the system.
It is seen that $\alpha$ converges to a value of 111.2$^{\circ}$
(Fig.~\ref{geom}b), consistent with the sp$^3$ bond angle (109.5$^{\circ}$),
while $\beta$ tends to a value of about 119.8$^{\circ}$, 
which is close to the ideal sp$^2$ bond angle (120$^{\circ}$).\cite{saito:98}
It should be noted that the converged values of the bond angles
coincide with the corresponding values in periodic (5,0) CNT.\cite{machon:02,liu:02}
The appearance of sp$^3$ hybridization in the systems 
comes from the significant curvature of the walls in such ultrathin CNTs.\cite{gulseren:02}
In order to provide more insights into the curvature effects in
(5,0) FCNTs, the highest occupied molecular orbital (HOMO), bond lengths, 
and bond angles of this system is compared with (7,0) and (9,0) finite tubes 
with the same length (table~\ref{curv}). 
The diameter of (5,0), (7,0), and (9,0) CNTs are about
3.91, 5.48, and 7.05\,\AA, respectively. 
In the (9,0) tube, which has the largest diameter,
the HOMO is seen to be localized on the edge atoms,
while the wall curvature of smaller tubes pushes this orbital
toward the central part of the system. 
Moreover, the bond lengths and bond angles of this system 
are close to the corresponding parameters in the ideal 
graphene (table~\ref{curv}), revealing the dominance of 
sp$^2$ hybridization in this system.
In the smaller finite tubes, the curvature effects increase
deviation of the bond parameters from the ideal graphene
toward diamond, and thus enhance sp$^3$ hybridization in the systems.

\begin{table}
\caption {\label{curv} 
 Comparing the highest occupied molecular orbital (HOMO)
 and structural parameters of three finite zigzag CNTs 
 with the same lengths and different diameters.
 L, T, $\alpha$, and $\beta$ have the same meaning as in Fig.~\ref{geom}
 and L$_0$ and $\alpha_0$ are equilibrium bond length and bond angle
 in ideal graphene.\cite{saito:98}}
\begin{ruledtabular}
\begin{tabular}{cm{2.3cm}cccc}
 FCNT &     ~~~~~HOMO  & L/L$_0$ & T/L$_0$ & $\alpha/\alpha_0$& $\beta/\alpha_0$ \\
\hline
(5,0) &\includegraphics[scale=1.5]{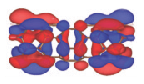} & 0.980 & 1.013 & 0.930 & 0.998 \\
(7,0) &\includegraphics[scale=1.5]{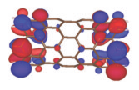} & 0.993 & 1.000 & 0.962 & 1.000 \\
(9,0) &\includegraphics[scale=1.5]{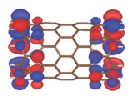} & 0.997 & 1.000 & 0.976 & 1.000 \\
\end{tabular}
\end{ruledtabular}
\end{table}

In order to see the effects of passivation on structural properties
of finite (5,0) CNTs, we tried to re-calculate some of the FCNTs 
in their pristine structure with no adsorbed hydrogen atoms. 
The presence of dangling bonds induces many difficulties in
the self-consistent calculation of these unpassivated systems.
After considerable efforts, we could only converge the self consistent
electronic structure of S2, S3, and S4.
It was found that, in contrast to the passivated CNTs, 
L bond becomes longer than T bond in these pristine FCNTs.
It means that the unpassivated FCNTs have slightly longer length and narrower diameter,
compared with the passivated ones.
This slight elongation may be attributed to the non-bonded electrons of the pristine FCNTs 
that induce an effective repulsive force between the two ends of the system.

We observe that the size dependency of the structural parameters,
presented in Fig.~\ref{geom}b, becomes negligibly small for tubes longer than S16.
Therefore, one may omit the costly process of structural relaxation
of longer (5,0) finite length carbon nanotubes by constructing
their structure from the relaxed atomic configuration of S16.
In order to verify this statement, we constructed the S18 tube from 
the structural parameters of S16 and then calculated its total energy and energy gap 
without and with a further structural relaxation.
The same calculations were performed for S20 and S22 tubes and the 
results were listed in table~\ref{error}. 
It is clearly seen that the finite tubes longer than S16, 
constructed from the atomic configuration of smaller tubes, 
in practice need not any further structural relaxation for investigating
their structural and electronic properties.

\begin{table}
\caption{\label{error} 
 Effect of a further structural relaxation on the total energy ($\Delta$E$_{tot}$)
 and energy gap ($\Delta$gap) of a longer finite CNT constructed from 
 the relaxed structure of the shorter tube. 
 }
\begin{ruledtabular}
\begin{tabular}{lccccc}
                      &  S14  &   S16  &   S18  &   S20  &   S22  \\
\hline
$\Delta$E$_{tot}$(Ry) & 0.360 &  0.090 &  0.009 &  0.002 &  0.003 \\ 
$\Delta$gap (Ry)      & 0.228 &  0.004 &  0.023 &  0.004 &  0.005 \\
\end{tabular}
\end{ruledtabular}
\end{table}

\begin{figure}
\includegraphics*[scale=0.85]{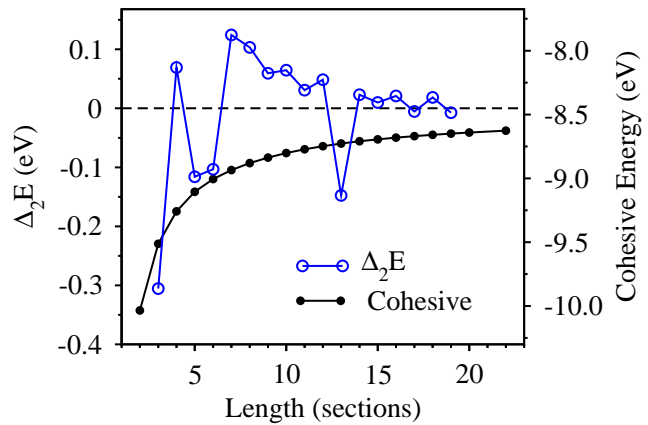}
\caption {\label{magic}
 Calculated cohesive energy (per carbon atom)  
 and second energy difference ($\Delta_{2}E$) of the (5,0) FCNTs. 
 The dashed line shows the cohesive energy of the periodic tube.
}
\end{figure}

The cohesive energy of the systems as 
a function of the tube length is shown in Fig.~\ref{magic}.
It is expected that by increasing the tube length, 
the cohesive energy converges to the periodic limit (8.50 eV/atom).
The absolute cohesive energy decreases as the length increases, 
indicating more stability of shorter tubes. 
It is attributed to the higher strength of C-H bond, compared with C-C bond.
The shorter FCNTs have higher fractional number of C-H bonds and hence are more stable.
Calculated harmonic vibrational frequencies of S2 and S3 provide
further evidence for this statement; there is a very strong peak
at about 3000\,cm$^{-1}$ which is related to the C-H bond. 
Barone \etal\ observed similar behavior in graphene nanoribbons.\cite{barone:06} 
They discussed that the stability of the passivated nanoribbons
is proportional to the molar fraction of hydrogen atom and
therefore inversely proportional to the width of the system.
In our case, the cohesive energy of FCNTs is inversely proportional 
to the tube length. 

\begin{figure*}
\begin{tabular}{ccccccccccccccc}
 \hspace{1.0cm} &
\includegraphics*[scale=0.120]{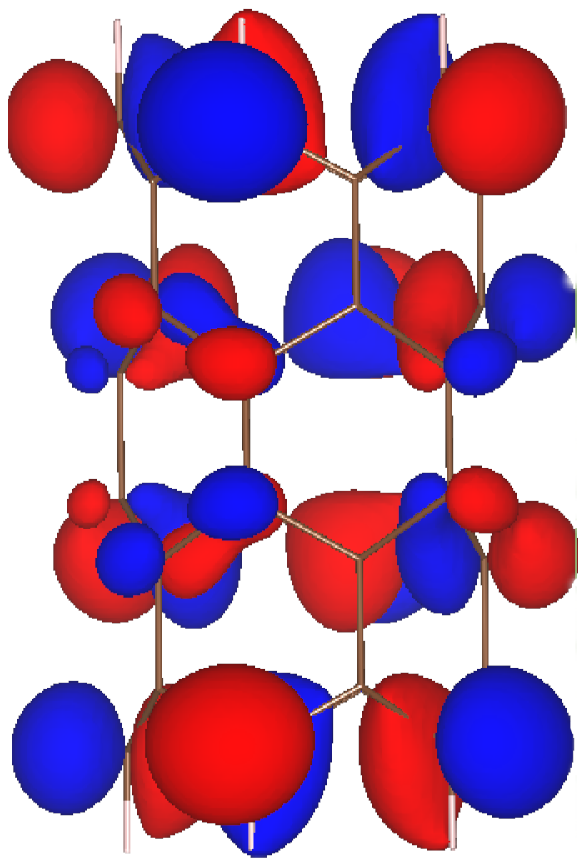} &
\includegraphics*[scale=0.126]{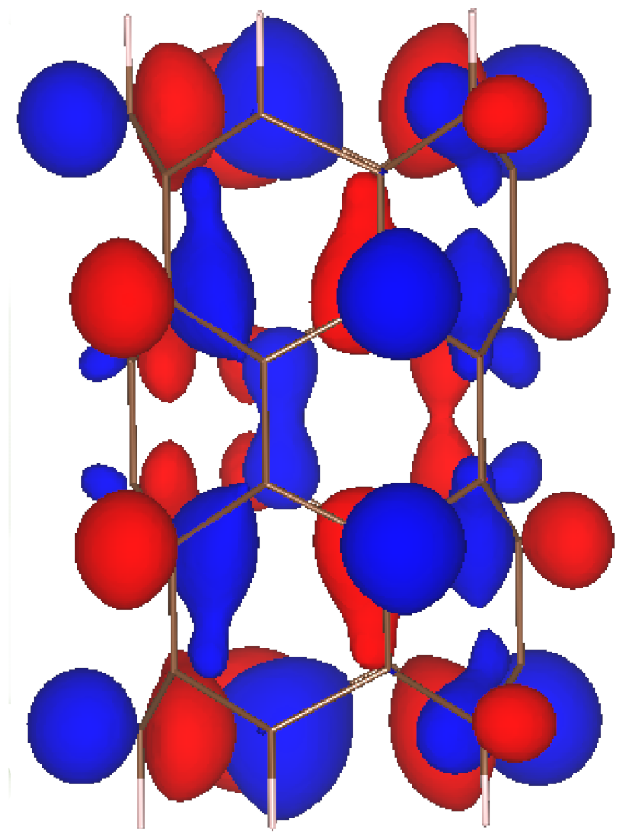} & \hspace{1.0cm} &
\includegraphics*[scale=0.096]{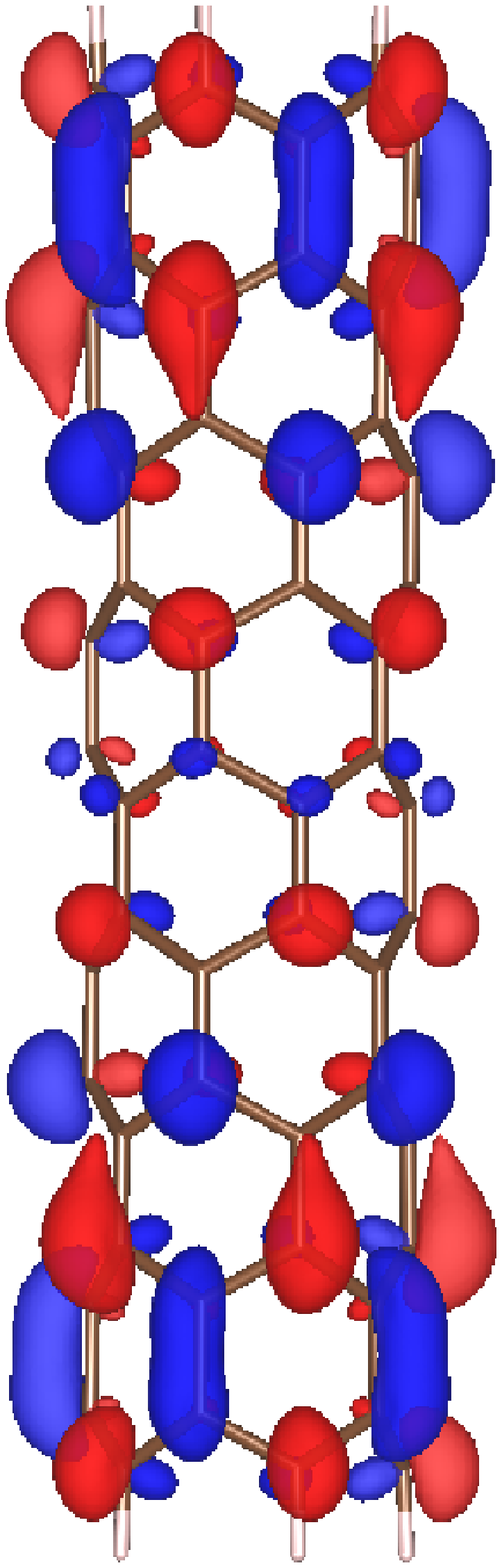} &
\includegraphics*[scale=0.096]{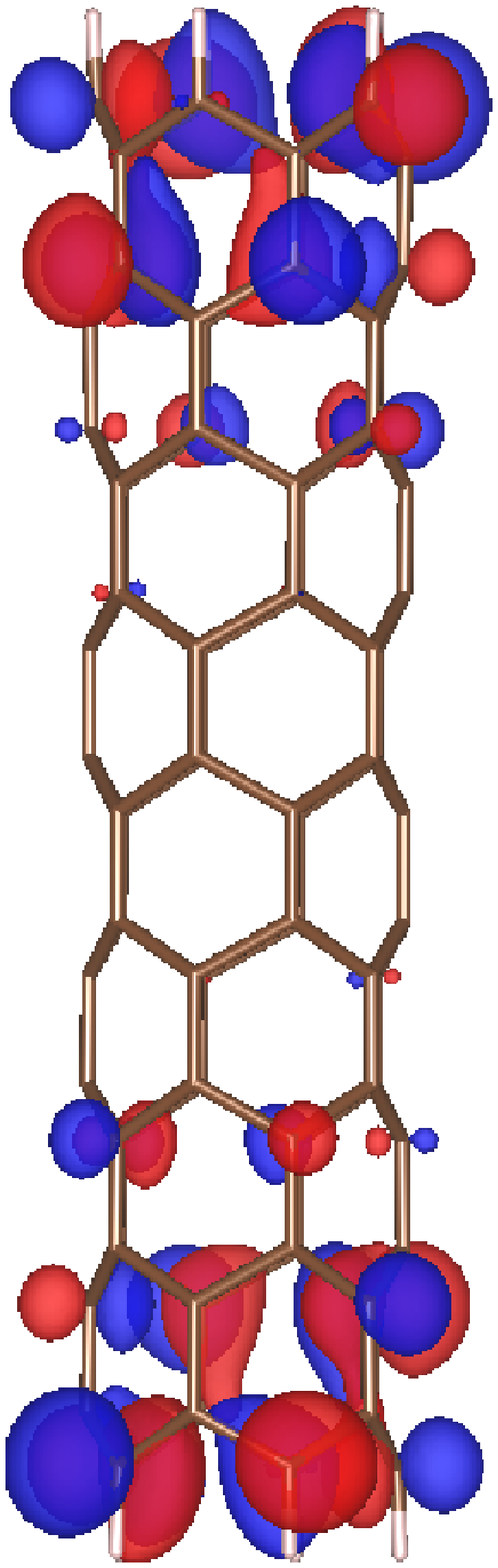} & \hspace{1.0cm} &
\includegraphics*[scale=0.120]{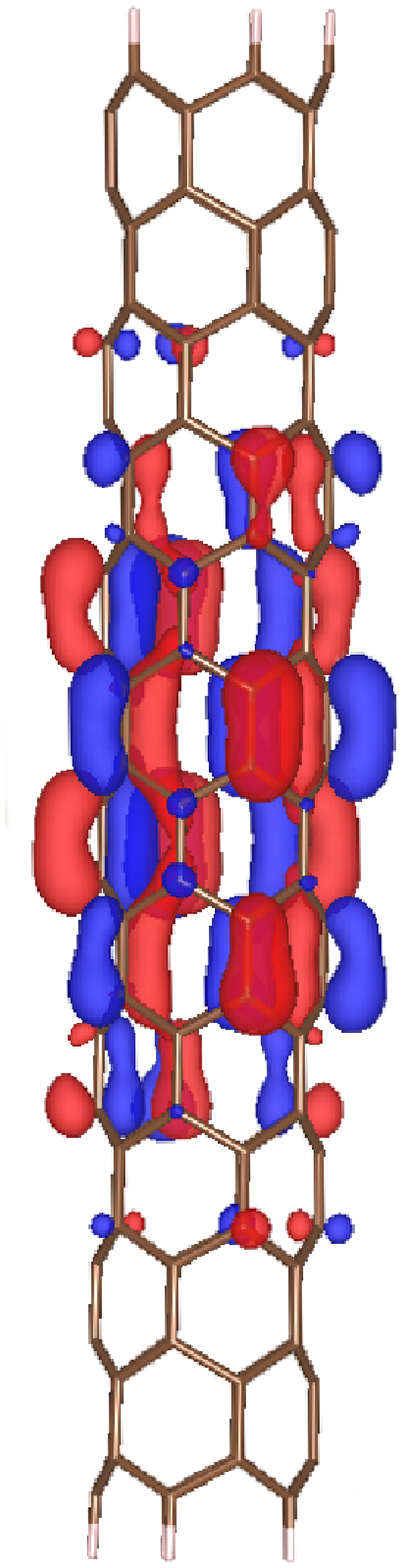} &
\includegraphics*[scale=0.126]{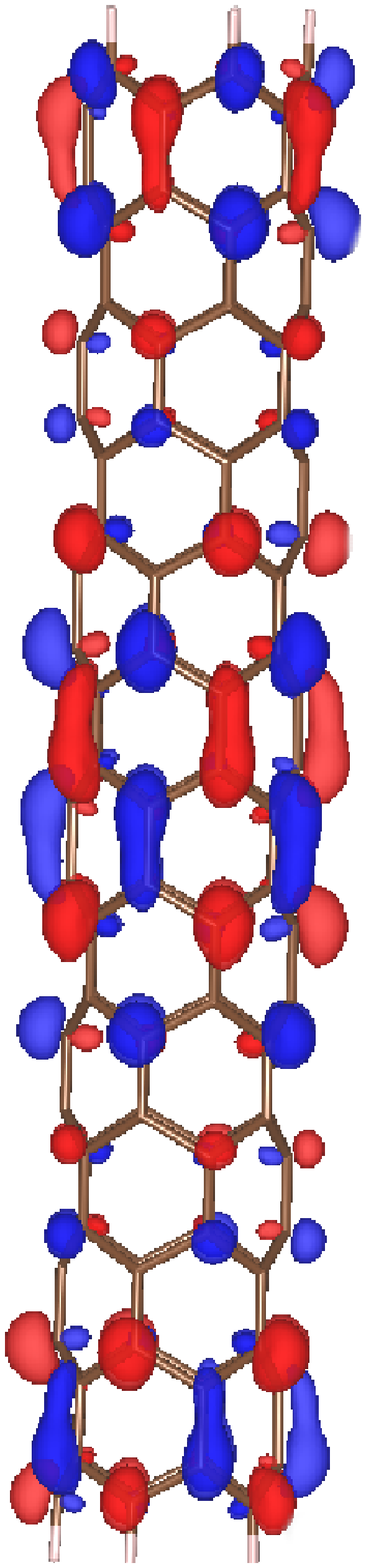} & \hspace{1.0cm} &
\includegraphics*[scale=0.114]{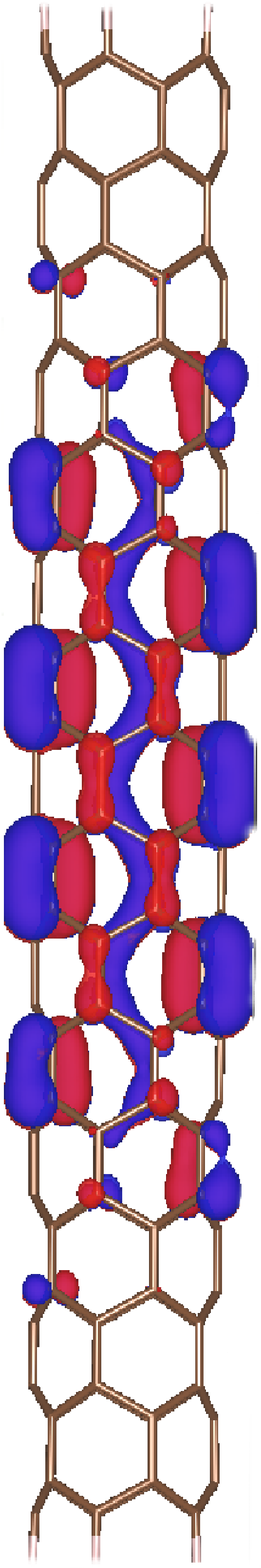} &
\includegraphics*[scale=0.108]{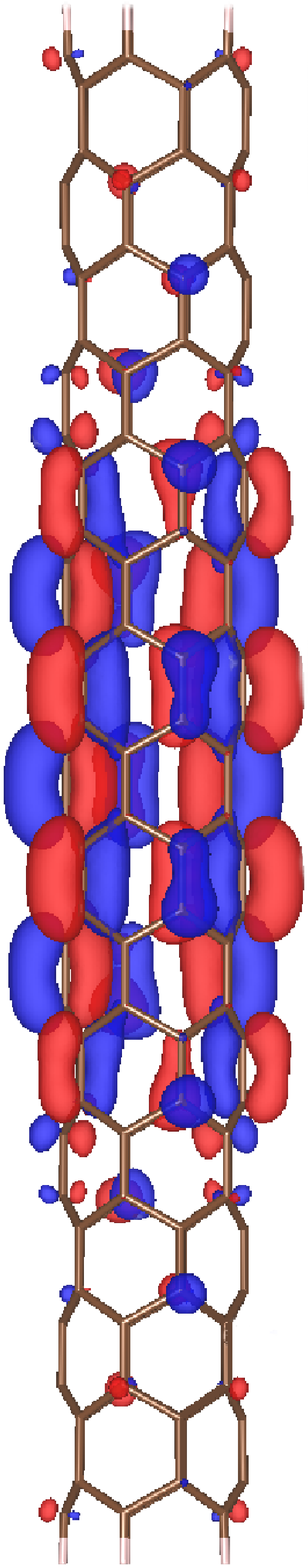} & \hspace{1.0cm} & 
\includegraphics*[scale=0.140]{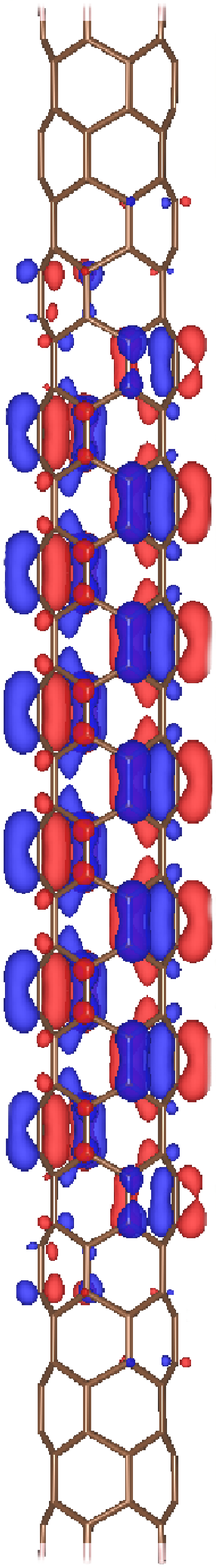} &
\includegraphics*[scale=0.146]{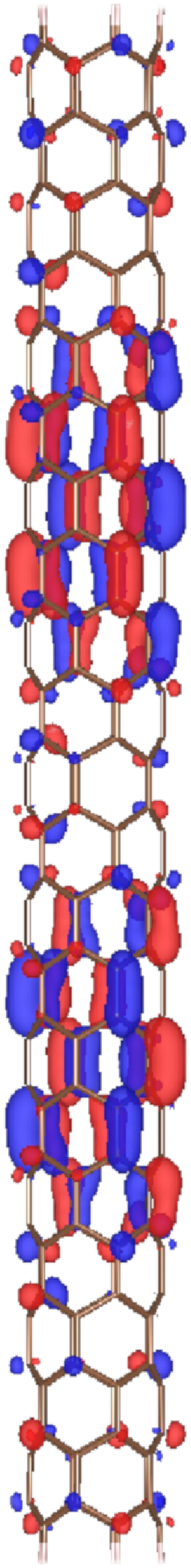} \\
 & ~S4-H~ & ~S4-L~ & & ~S9-H~ & ~S9-L~ & & ~S13-H~ & ~S13-L~ & &
 ~S16-H~ & ~S16-L~ & & ~S22-H~ & ~S22-L~ \\ 
\end{tabular}
\includegraphics*[scale=0.90]{fig3-gap}
\caption { \label{gap}
 Top: Isosurfaces of the highest occupied molecular orbital (HOMO) and the lowest unoccupied 
 molecular orbital (LUMO) of (5,0) FCNTs with different length.
 Bottom: Calculated energy gap of the finite tubes as a function of the tube length. 
 Inset: the electronic band structure of a graphene nanoribbon with a
 width of four sections.}
\end{figure*}

On the other hand, Rochefort \etal\ observed a reverse behavior,\cite{rochefort:99} 
their absolute cohesive energy of CNTs per atom increases as the length increases. 
This difference may be ascribed to the different energy reference used
for calculation of cohesive energy in this work.
Following some benchmark calculations, we realized that using a reference energy 
which includes C-H bond, like benzene or methane, give rises 
to a cohesive energy which decreases (become more negative) by increasing length.
These statements may be helpful to identify the required experimental
setup for synthesis of short or long FCNTs. 
It seems that in a high vacuum environment, long FCNTs are more feasible
while production of short FCNTs may be more practical in the presence of
a proper hydrogen gas flux.

In the study of atomic clusters, a conventional parameter for addressing 
relative stabilities is the second-order difference of total energy
($\Delta_{2}E$), defined as:
\begin{equation} \label{del2}
\Delta_{2}E(i)=E_{tot}(i+1)+E_{tot}(i-1)-2E_{tot}(i)
\end{equation}
where $E_{tot}(i)$ denotes to the relaxed total energy of the 
cluster of size $i$. 
The positive values of this parameter determine the so called
magic sizes of the system which exhibit higher stability with respect 
to the adjacent sizes. 
The second-order energy difference of the given FCNTs
(Fig.~\ref{magic}) indicates that S7 and S8 are energetically
the most favorable sizes, while S3, S5, S6, and S13 show 
the least relative stabilities. 
This plot is expected to be comparable with the envisaged 
mass spectra measurements on (5,0) FCNTs. 

\section{Electronic properties}

The calculated energy gap of the finite CNTs, in the non-magnetic state,
as a function of their length is represented in Fig.~\ref{gap}. 
In order to understand this behavior, one has to take into account
the zone-folding and the curvature effects.
The zone-folding approach is a popular method to estimate
the electronic structure of a nanotube from its graphene counterpart.
This approach predicts a semiconducting gap
for infinite (3n$\pm$1,0) zigzag CNTs,\cite{hamada:92} 
which in all cases larger than (5,0) CNT, agrees with the accurate DFT based calculations.
The metallic behavior of the (5,0) CNT 
is attributed to the high curvature of this narrow tube,
which is not taken into account in the zone-folding approach.\cite{gulseren:02,popov:04}
The high curvature of this system enhances the interatomic interactions
and thus increases the dispersion of valence and conduction bands.
Consequently, these bands overlap and form a metallic system.
We will discuss that the curvature effects may have quite different impacts 
in the zone-folded electronic structure of the finite CNTs,
which is derived from the electronic structure of graphene nano-ribbons (GNRs).

The electronic band structure of the GNR corresponding to S4, is calculated
and presented in the inset of Fig.~\ref{gap}.
It is seen that the valence and conduction bands are degenerate and flat
near edge of the Brillouin zone (BZ).
The width of the degenerate region increases by increasing the width of the GNR.
Folding the GNR will discretize the allowed k-points of a (n,0) FCNT as:
$k_{N}=\frac{2\pi}{a}\frac{N}{n}$ $(N=0,\pm1,\pm2, \cdots)$
where $a$ is the GNR lattice parameter.\cite{okada:03} 
For a (5,0) tube, five k-points satisfy the boundary condition,
in which the last one falls on the degenerate region and leaves a zero energy gap 
in the folded electronic structure of the systems.
However, the obtained results (Fig.~\ref{gap}) indicate that 
all FCNTs, except S3 and S4, exhibit a non-trivial energy gap.
The reason is that the high curvature of these finite systems enhances 
the interatomic interactions and consequently 
increases dispersion of the discrete energy levels
to open a gap between the highest occupied (HOMO) and the lowest unoccupied (LUMO) levels.
In other words, the curvature effects act oppositely in the finite and periodic CNTs;
increasing band dispersion in a periodic CNT may give rise 
to overlapping of the valence and conduction bands, whereas increasing dispersion
of discrete energy levels of a finite CNT may enhance the HOMO-LUMO splitting.

We use the difference between the average L and T bond lengths ($\Delta$bond), 
as a proper measure of the curvature of an FCNT.
It is rationalized by the fact that this difference
decreases toward zero by increasing the CNT diameter (table~\ref{curv}).
The results, presented in Fig.~\ref{geom}, 
indicate that in the smallest FCNT (S2), L bond is on average longer than T bond.
By increasing the FCNT length, the L bond length falls quickly while 
the T bond length rises sharply.
It is clearly observed that within S3 and S4 the longitudinal
and transverse bond lengths are getting very close together.
As a result, these systems involve the lowest curvature effects and hence
their zero energy gap in the non-magnetic state agrees with the zone-folding prediction.
On the other hand, S2 exhibits the highest curvature effect and 
consequently a significant energy gap appears in this system.
We observe that from S4 to about S9, $\Delta$bond and consequently curvature are growing.
This trend may explain the increment of energy gap between S4 and S9 (Fig.~\ref{gap}).
For FCNTs longer than S9, $\Delta$bond is almost converged to the periodic CNT value
and hence the curvature is saturated in this region.
Therefore, the asymptotic decrease of the energy gap in this region should 
be explained by a different effect.

In larger FCNTs with lower curvature effect, 
similar asymptotic behavior, observed in a broad length range, 
is attributed to quantum confinement effects.\cite{zhu:98}
In the simple picture of {\em particle in a one-dimensional box},
the distance between quantum energy levels decreases asymptotically
with respect to the box length.
We observe that in the short (5,0) FCNTs, significant variation of curvature 
destroys the asymptotic behavior of the energy gap,
while after stabilization of the curvature around S9, 
the asymptotic decrease of the quantum confinement appears (Fig.~\ref{gap}).
However, the results show the decrease of energy gap stops around S15 
and converges to a value of about 0.2\,eV afterwards,
whereas, the periodic (5,0) CNT has a metallic character within DFT.
Therefore, long-range edge effects are expected in the electronic
and optical spectra of (5,0) finite length CNTs.
For better understanding of this behavior, 
the HOMO and LUMO isosurfaces of some investigated FCNTs 
are presented in the top of Fig.~\ref{gap}.
It is apparent that the size dependency of HOMO is limited to the tubes 
shorter than S10, while for longer tubes this orbital exhibits a uniform
distribution over the central sections of the FCNT.
In contrast, LUMO exhibits persistent size dependent fluctuations
even up to the size of S22 (44\,\AA).
This observation provides more evidence for the resistant edge effects 
in the excited state properties of (5,0) FCNTs.

As it was mentioned earlier, the curvature of S3 and S4 is
compensated and hence the zero energy gap of these systems is mainly
derived from the zone-folding of the GNR sp$^2$ band structure.
In other investigated FCNTs, the curvature effects play a significant role and 
induce a non-trivial sp$^3$ hybridization in the systems.
In the case of sp$^2$ hybridization, the system has equivalent L and T bonds,
just as graphene, while contribution of sp$^3$ hybridization in the systems
with high curvature increases the length of the transverse T bond, 
compared with the longitudinal L bond, and opens a gap in the systems.
This statement is justified by the behavior of $\Delta$bond in Fig.~\ref{geom}.

\begin{table}
\caption{\label{mag} 
  Calculated magnetic energy ($\Delta$E), spin moment ($\mu$),
  and spin density isosurfaces of some (5,0) FCNTs in the ferromagnetic state.
  The numbers in the parentheses show magnetic energy of the systems
  in the antiferromagnetic state.
  }
\begin{ruledtabular}
\begin{tabular}{lccm{2.0cm}}
FCNT   & $\Delta$E (meV) & $\mu$ ($\mu_B$) &   spin density       \\
\hline\vspace{-0.1cm}
       &           &      &     \\ 
 S3    & -130 (0)  & 2.00 & \includegraphics[scale=0.20]{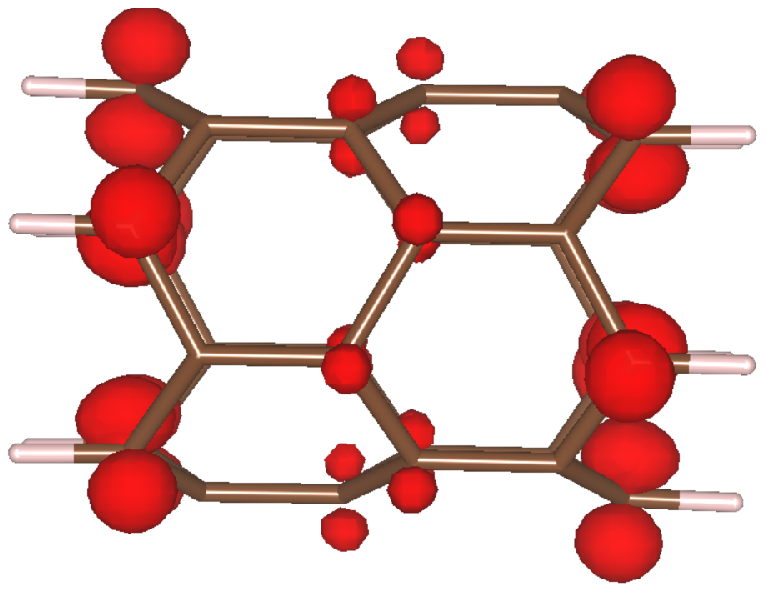} \\ 
 S4    & -127 (-6) & 2.00 & \includegraphics[scale=0.25]{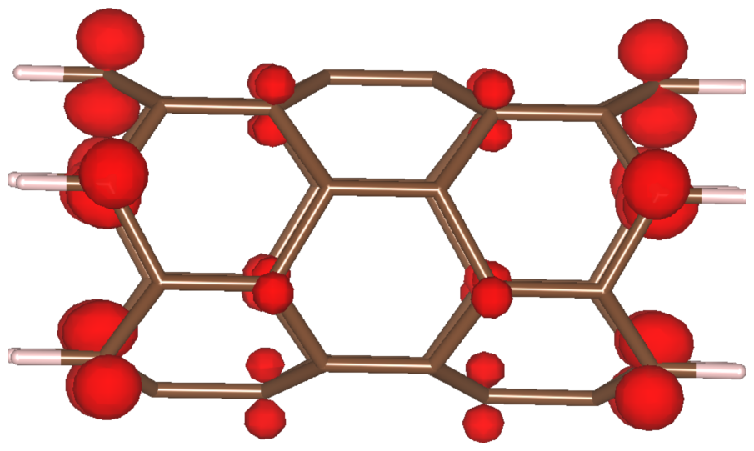} \\
 S5    &   0  (2)  & 2.00 & \includegraphics[scale=0.30]{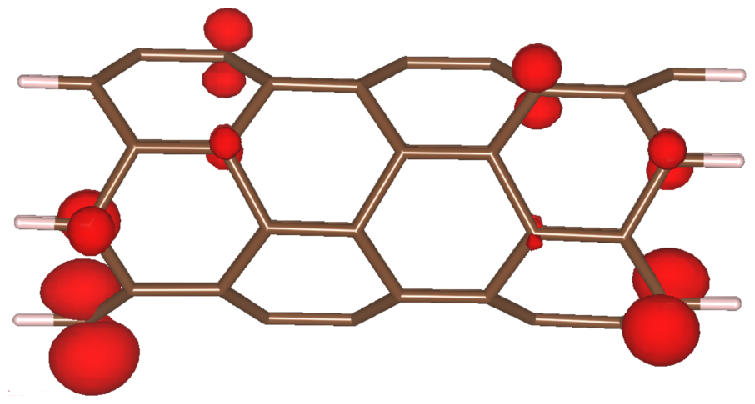} \\
\end{tabular}
\end{ruledtabular}
\end{table}

The zero energy gap of S3 and S4 zigzag FCNTs shows
the potential of spontaneous spin polarization in the system.
From a general point of view, the zigzag graphitic materials are predicted 
to possess a ferromagnetic (FM) or antiferromagnetic (AF) spin-polarized 
ground state.\cite{lieb:89} 
Therefore, we studied the FM and AF states of the considered FCNTs 
within spin-polarized calculations.
It was found that S3, S4, and S5 have FM and AF states (table~\ref{mag}), 
while all other FCNTs exhibit no spontaneous spin polarization.
The calculated spin densities reveal that the magnetic moment 
of S3, S4, and S5 is concentrated on the edge carbon atoms.
The total spin moment of the spin-polarized systems is found to be 2\,$\mu_B$,
because there are two allowed k-points ($\pm4\pi/5a$) in the degenerate region of 
the corresponding GNR band structure.

The magnetic energy of S3, S4, and S5 is also presented in table~\ref{mag}.
This parameter measures the difference between the minimized energy
of system in the non-magnetic (NM) and corresponding magnetic states.
We observe that S3 and S4 prefer a FM state which
is significantly more stable than the NM and AF states.
In the case of S5, the FM, AF, and NM state have very close minimized energies
while the FM state is marginally more stable.
We found that FM spin polarization has significant effect on the electronic
structure of S3, S4, and S5.
As it was expected, spin polarization removes the degeneracy of the HOMO
and LUMO levels in S3 and S4 and opens a gap in these systems.
The enhanced energy gap of these systems in the FM state is presented in Fig.~\ref{gap}.
It should be noted that the minimized total energy of S3, S4, and S5,
presented in Fig.~\ref{magic} are determined in the FM state.

\begin{figure}
\includegraphics*[scale=0.9]{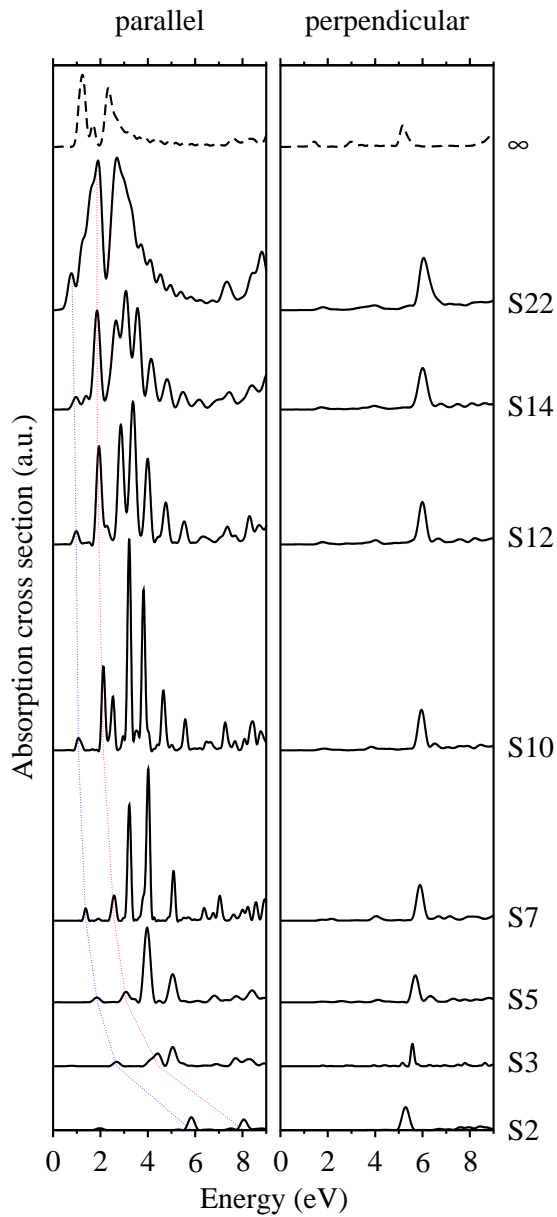}
\caption {\label{sig}
 Absorption cross section of the investigated FCNTs. 
 The tube lengths are indicated in the right axis.
 The symbol $\infty$ stands for an infinite (periodic) (5,0) CNT.
 The evolution of the two first absorption peaks of the FCNTs
 are highlighted by the dotted and dashed lines. 
}
\end{figure}

\section{Optical spectra}

As it was mentioned, we used time dependent DFT to calculate 
absorption spectra of the desired finite CNTs in the presence of
a polarized external electromagnetic field. 
The calculated absorption cross sections, for parallel and perpendicular polarization,
are presented in Fig.~\ref{sig}.
In the parallel polarization, the shorter tubes exhibit more sparse spectrum, 
originating from their discrete energy levels,
while growing the tube length, increases the number of peaks and 
accumulates them together.
It is due to the fact that by growing the tube length, the discrete energy states
are getting more dense and gradually transforming to one dimensional bands.
We observe that decreasing the tube length quenches the absorption intensities.
This quenching is attributed to the local field effects (LFEs) 
or depolarization field emerged from microscopic electric dipoles 
induced by the external field.
Fortunately, LFEs are well treated in the TDDFT formalism.\cite{marinopoulos:03} 
It is argued that the strength of LFEs decreases rapidly 
by increasing the relevant dimension of the system.\cite{spataru:04}
As a result, it is seen that growing the tube length significantly amplifies
the absorption spectra in the parallel polarization,
while fixed lateral dimension of FCNTs reduces the size sensitivity of 
the perpendicular absorption spectra. 
This lateral confinement give rises to strong perpendicular LFEs 
and hence quenches the absorption peaks in this polarization. 

\begin{figure}
\includegraphics*[scale=0.4]{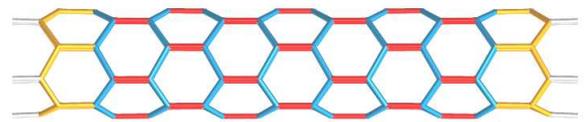}
\caption {\label{edge}
 Bond pattern of S10;
 red bonds are shorter than 1.40\,\AA,
 yellow bonds are between 1.40 and 1.44\,\AA,
 and blue bonds are longer than 1.44\,\AA.
}
\end{figure}

For better understanding of the absorption spectra in the parallel polarization, 
the behavior of the first two absorption peaks are highlighted by using 
two dashed lines in Fig.~\ref{sig}.
The first dashed line (blue), which likely mimics trend of the optical gap
of FCNTs, exhibits an obvious red shift, more pronounced in the shorter tubes.
In other words, the optical gap of these systems is rapidly decreasing
from S2 to S7 and then converging to the obtained optical band gap 
of periodic (5,0) tube ($\sim$1.2\,eV).
In contrast to our results, Montero \etal\,\cite{montero:10} 
observed a blue shift in the optical absorption of (5,0) FCNTs,
computed using post Hartree-Fock method.
Accurate inspection of their published paper shows that
they have likely constructed their finite tubes by repeating 
the relaxed units of the periodic tube and no further structural 
relaxation has been applied to the finite systems.
Our calculations indicate significant relaxation effects on edge atoms
which affects the optical spectra of finite systems, specially for shorter tubes.
The bond pattern of S10, as a prototype, is presented in Fig.~\ref{edge}.

Compared with the electronic gap, presented in Fig.~\ref{gap},
optical gaps are significantly larger and display different trend.
It may evidence absence of any HOMO-LUMO optical transition in the (5,0) FCNTs.
In order to verify this statement, we used the linear response Casida approach 
to characterize the first allowed optical transitions of four FCNTs. 
The results, given in table~\ref{casida},
confirm absence of the HOMO-LUMO transition in the systems.
In the case of S3 and S10, we observe that the first optical transition 
in the parallel polarization occurs from the second level below HOMO to 
the fourth and fifth levels above LUMO, respectively.
In the case of S5, in the parallel polarization, 
and S7 and S10, in the perpendicular polarization, more than one joint levels 
was found to contribute to the first allowed optical transition.
In the periodic CNTs, it is already discussed that 
the first allowed transition, in the parallel polarization,
occurs between the valence and conduction bands 
which have identical orbital character.\cite{reich:08}
Our results show that lack of translational symmetry changes
the selection rules and hence prevents the HOMO-LUMO transition
in the finite CNTs.

\begin{table}
\caption{\label{casida}
 Character of the first allowed optical transition of S3, S5, S7, and S10
 in the parallel and perpendicular polarizations.
 The capital letters L and H stand for LUMO and HOMO.
  }
\begin{ruledtabular}
\begin{tabular}{lclclclclcl}
FCNT   &&\multicolumn{3}{c}{parallel} &&\multicolumn{3}{c}{perpendicular}\\
\hline
 S3    && H$-2$ &$\rightarrow$& L$+4$ && H$-15$    &$\rightarrow$& L$+5$  \\
       &~~~~~~~~&&            &       &&        &             &         \\
 S5    && H$-1$ &$\rightarrow$& L$+1$ && H$-6$    &$\rightarrow$& L$+25$  \\
       && H     &$\rightarrow$& L$+4$ &&                                  \\
       &~~~~~~~~&&            &       &&        &             &         \\
 S7    && H     &$\rightarrow$& L$+6$ && H$-9$ &$\rightarrow$& L$+20$   \\
       &~~~~~~~~~&&           &       && H$-10$ &$\rightarrow$& L$+19$   \\
       &~~~~~~~~&&             &       &&        &             &        \\
 S10   && H$-2$ &$\rightarrow$& L$+5$ && H$-39$ &$\rightarrow$& L$+14$   \\
       &~~~~~~~~~&&           &       && H$-40$ &$\rightarrow$& L$+13$  \\
       &~~~~~~~~~&&           &       && H$-42$ &$\rightarrow$& L$+16$  \\
       &~~~~~~~~~&&           &       && H$-43$ &$\rightarrow$& L$+15$  \\
\end{tabular}
\end{ruledtabular}
\end{table}

\begin{figure}
\includegraphics*[scale=0.9]{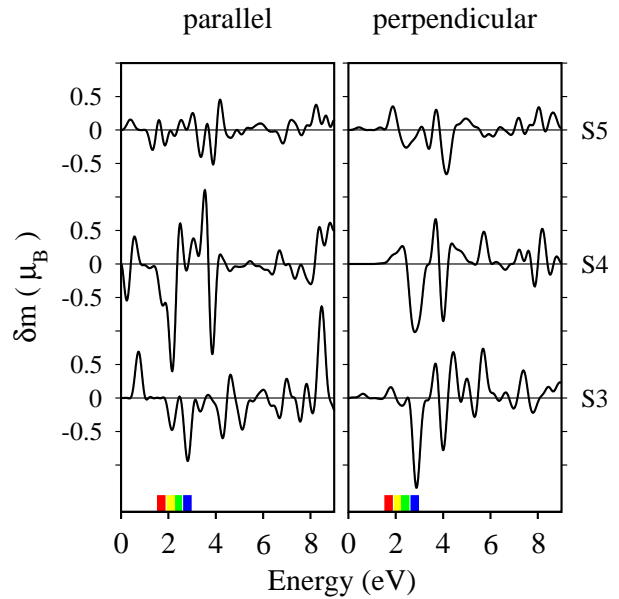}
\caption {\label{dyn-mag} 
Calculated dynamical spin dipole of S3, S4, and, S5.
$\delta\,m(\omega)=\delta\,n^{\uparrow}(\omega)-\delta\,n^{\downarrow}(\omega)$
where $\delta\,n^{\uparrow(\downarrow)}(\omega)$ denotes the dynamical spin
response of the system in the majority (minority) channel.\cite{torres:00}
The energy range of the visible spectra is highlighted on the energy axes.
}
\end{figure}

In the systems with spin polarized electronic structure, one may expect some 
magnetic responses to the photon (electromagnetic) irradiation, 
while non-magnetic systems exhibit only electric response to an  
external electromagnetic field.\cite{yin:09}
Therefore, we calculated the spin dipole response of S3, S4, 
and S5 FCNTs to an external magnetic field, in the framework of TD-LSDA. 
The obtained spectra, presented in Fig.~\ref{dyn-mag},
illustrate stronger magnetic response of S3 and S4,
likely due to their higher ferromagnetic energy, compared with S5 (table~\ref{mag}).
It is interesting to see that the higher responses of S3 and S4 happens
in the visible spectrum.
Comparing magnetic responses of the systems in the parallel and 
perpendicular polarizations shows that, in contrast to the electric responses, 
the effects of magnetic LFEs are not evident.
The reason is that the magnetic moment of the systems, as it was mentioned earlier,
is rather distributed over the edge carbon atoms (see table~\ref{mag}).
As a result, the magnetic response mainly comes from the edge atoms and
hence is less sensitive to the polarization.
The interplay between optical excitations and dynamical magnetizations
of FCNTs is promising to develop new nano devices for magneto-optical opportunities.

\section{Conclusions}

In summary, full-potential density functional computations was employed to 
investigate structural and electronic properties of (5,0) finite
carbon nanotubes (FCNTs), passivated with hydrogen atoms.
The behavior of bond lengths and angles as a function of the tube length
indicated that the ground state properties converge to the periodic limit
at a length of about 30\,\AA.
We discussed that the zone-folded electronic structure of the finite CNTs,
in contrast to the periodic system, shows no energy gap,
while high curvature of (5,0) FCNTs opens a gap in the system.
It was argued that compensation of the curvature effects at a length
of about 6-8\,\AA, preserves the sp$^2$ character of carbon-carbon bonds
and closes the energy gap of FCNT in the nonmagnetic state,
and then brings in a ferromagnetic state.
Analysis of the second-order differences of total energy showed that for 
the (5,0) FCNTs shorter than 40\,\AA, a length of about 15\,\AA
exhibits the highest relative stability.
The behavior of the energy gap and the lowest unoccupied molecular orbital indicated that,
in contrast to the ground state properties, 
the excited state properties of the systems exhibits long range edge effects.
The real time propagation of Kohn-Sham orbitals was applied to determine
the electric dipole and spin dipole responses of FCNTs in the presence 
of electromagnetic pulses. 
The calculated optical absorption spectra show that the optical gap of 
the systems is considerably larger than the Kohn-Sham energy gap.
The obtained character of optical transitions in the framework of Casida equation 
confirms absence of any optical transition between the highest unoccupied and 
the lowest unoccupied molecular orbitals in (5,0) finite CNTs. 
The spin dipole response of the systems,
which is found to be more pronounced in the visible or near visible region,
is of order of Bohr magneton.

\section{Acknowledgments}
We acknowledge the helpful comments of Mojtaba Alaei 
(Isfahan University of Technology).

\end{document}